\documentclass{jnmp}

\JNMPnumberwithin{equation}{section}
\usepackage{amsmath}

\begin{document}

\renewcommand{\evenhead}{H Leblond}
\renewcommand{\oddhead}{Mixed  Perturbative Expansion: the Validity
of a Model for the Cascading}

\thispagestyle{empty}

\FirstPageHead{9}{2}{2002}{\pageref{leblond-firstpage}--\pageref{leblond-lastpage}}{Article}

\copyrightnote{2002}{H Leblond}

\Name{Mixed  Perturbative Expansion:\\ the Validity
of a Model for the Cascading}\label{leblond-firstpage}

\Author{Herv\'e LEBLOND}

\Address{Laboratoire POMA, UMR CNRS 6136,
Universit\'e d'Angers,\\
2 ${\it B^d}$ Lavoisier 49045 ANGERS Cedex 01,
France\\
E-mail: herve.leblond@univ-angers.fr}

\Date{Received May 31, 2001; Revised September 17, 2001; Accepted
December 11, 2001}

\begin{abstract}
\noindent
A new type of perturbative expansion is built in order to give a
rigorous derivation and to clarify the range of validity of some
commonly used model
equations.
This model describes the evolution of the modulation of two short and
localized pulses, fundamental and second harmonic, propagating together in a
bulk uniaxial crystal with non-vanishing second order susceptibility $\chi^{(2)}$,
 and interacting through the nonlinear effect known as ``cascading'' in nonlinear optics.
  The perturbative method mixes a~multi-scale
expansion with a power series expansion of the susceptibility, and must be
carefully adapted to the physical situation. It allows the determination of
the physical conditions under which the model is valid: the order of
magnitude of the walk-off, phase-mismatch, and anisotropy must have
determined values.
\end{abstract}

\section{Introduction}

\subsection{A methodological problem}

Many scientists, in recent times, came to the idea that physics is
intrinsically nonlinear and that most of the linear theories have been
introduced because they were mathematically tractable, contrarily to the
nonlinear problems. Nowadays, the theoreticians of nonlinear physics almost
always try to keep the largest part of the ideas involved by the linear
theories of the  physical problems considered. Both mathematical tools and
physical concepts are conserved. The nonlinear theory appears then mainly as
a perturbative approach in which the linear theory is the zero order. As an
essential example, the Fourier theory and the plane monochromatic wave
concept are the main tools in the linear theory of wave propagation. The
theory of nonlinear wave propagation considers some wave packet, that is in
fact a monochromatic plane wave (i.e.\ the linear wave) modulated by some
slowly varying envelope (which is thus some first correction to the linear
theory). This theory involves thus at least two space scales: that of the
wavelength and that of the pulse length, which are treated separately by
the mathematical model. The natural frame for the derivation of the
nonlinear models is thus the so-called multiscale expansion method. From the
mathematical viewpoint, formal computations in this frame can be performed
in a perfectly rigorous way. Convergence proofs have also been given~\cite{colin}.
They enhance and clarify the validity of the formal computations.

A multiscale expansion, as well as any asymptotic expansion, has a
structural drawback. It involves some small parameter $\varepsilon
$, that is infinitely small in the mathematical theory, while it
takes a small but finite value in the physical applications.
Consider then some physical quantity, small in relation to the
physical value of $\varepsilon $, but independent of it. Because
it is finite while  $\varepsilon $ is not, the
mathematical theory considers it {\it de facto} as if it were
large in relation to $\varepsilon $. This may sometimes introduce
erroneous results, in so far that some further approximation is
not taken into account. To introduce, beside the slowly varying
envelope approximation and mixed to it, a further approximation of
another type yields great technical difficulties. Such an
approach mixing a multiscale expansion to an approximation of
another type has already been done for the treatment of damping
for soliton propagation, in particular for electromagnetic waves
in ferromagnets \cite{NakataII,amor}. The damping
parameter is identified to some power of the perturbative
parameter~$\varepsilon $, and all involved quantities are expanded
in a power series of the damping constant.

In other situations, the ansatz must be constructed through a
careful analysis of the physical situation. We study in this paper
a situation in which such a mixed expansion is needed. It is
presented from a methodological point of view, and should be of
interest for the readers who intend
 to apply the same kind of perturbative approach to other
physical situations. The proper physical consequences of the
result will be published elsewhere~\cite{match2}.

\subsection{Solitons through cascading}

The physical situation we consider is the propagation of optical
solitons using the so-called cascading phenomenon, viz.
second-harmonic generation and back-conversion to the fundamental,
that arises in a dielectric medium with a non-vanishing
second-order nonlinear susceptibility~$\chi ^{(2)}$. Several
theoretical results have been published in order to show that this
phenomenon could be responsible for the stabilization of a short
pulse and for solitonic behavior
\cite{berge,kanashov,menyuk,torner}. These theoretical works study
by analytic and numerical methods some partial differential
system, derived in the frame of the slowly varying envelope
approximation, and analogous in its form to the nonlinear
Schr\"{o}dinger (NLS) equation, at least for the linear part. The
nonlinear part of this system gives account for the interaction
between the fundamental and the second harmonic. Some applications
of this model have been found in good agreement with experimental
results. However, the way it is derived in the literature is quite
phenomenological. It appears thus to be worth giving a rigorous
formal derivation of that model, and to clarify its range of
validity.

Two waves propagating together are considered here, and their
interaction is expected to stabilize the pulse. Therefore two
conditions must be satisfied, at least in an approximate way.
First the interaction must be resonant, thus the waves must be
phase-matched. If not the interaction will not be strong enough to
yield the expected effect. Second the two waves must propagate
together during a long enough time. Thus their group
velo\-ci\-ties must be equal, or at least very close together. We
determine precisely the order of magnitude of the allowed small
deviation from phase-matching and small walk-off. We restrict for
technical reasons our attention to crystals having some particular
symmetry. However, the required phase and group velocity matching
conditions cannot be realized in any crystal of this type: the
linear dispersion relation of the material must satisfy several
conditions. The order of magnitude of some physical quantities
must be comparable to powers of the perturbative parameter
characteristic of the multiscale expansion.

\section{Analysis of the physical problem}

\subsection{The equations for a uniaxial optical crystal}

We consider a bulk crystal with a non-zero second order susceptibility
($\chi ^{(2)}$) tensor, optically uniaxial from the linear point of view and
perfectly transparent. These hypotheses are satisfied by potassium
dihydrogen phosphate (KDP) and analogous materials or by lithium niobate,
as examples. Wave propagation in such a medium is described by the Maxwell
equations that reduce to
\begin{equation}
\vec{\nabla}\wedge \left( \vec{\nabla}\wedge \vec{E}\right) =-\frac{1}{c^{2}}\,
\partial _{t}^{2}\left( \vec{E}+\vec{D}\right),
\end{equation}
where $c$ is the speed of light in vacuum, and
\begin{gather}
\vec{D}= \chi ^{(1)}*\vec{E}+\chi ^{(2)}*\vec{E}\vec{E} \nonumber\\
  \phantom{\vec{D}}= \int_{-\infty }^{t}\chi ^{(1)}(t-t^{\prime })\vec{E}
(t^{\prime })dt^{\prime }+\int_{-\infty }^{t}\int_{-\infty }^{t^{\prime
}}\chi ^{(2)}(t-t^{\prime },t-t^{\prime \prime }):\vec{E}(t^{\prime })\vec{E}
(t^{\prime \prime })dt^{\prime \prime }dt^{\prime }.
\label{de}
\end{gather}
$\chi ^{(1)}$ and $\chi ^{(2)}$ are respectively the linear and the
quadratic susceptibility tensors \cite{boyd}. We denote by $\hat{\chi}^{(1)}(\omega )$
 the Fourier transform of $\chi ^{(1)}(t)$ so that
$\chi^{(1)}(t)=\int_{-\infty }^{+\infty }\hat{\chi}^{(1)}(\omega )e^{i\omega
t}d\omega $. Then, for a uniaxial crystal, the coordinate frame being chosen
so that the optical axis is the $z$-axis, the linear susceptibility tensor
satisfies
\begin{equation}
1+\hat{\chi}^{(1)}=\left(
\begin{array}{ccc}
n_{o}^{2} & 0 & 0 \\
0 & n_{o}^{2} & 0 \\
0 & 0 & n_{e}^{2}
\end{array}
\right). \label{43}
\end{equation}
$n_{o}(\omega )$ and $n_{e}(\omega )$ are called respectively the ordinary
and extraordinary indices \cite{born}. A~wave propagates along the optical axis with the
velocity $c/n_{o}$.
 If the propagation direction makes
some nonzero angle $\theta $ with this axis, two waves can propagate, with different
speeds and polarizations. They are referred to as ordinary and extraordinary
waves. The characteristic feature of the extraordinary wave is that its group velocity
is not parallel to its phase velocity, which never happens in isotropic media.
If $\theta=\pi/2$, the wave polarized perpendicular to the optical axis is the ordinary one,
the wave polarized along the axis is extraordinary, and propagates with the velocity $c/n_{e}$.

\subsection{The phase-matching angle}

No wave interaction can occur far from the phase resonance, or
phase-matching. Therefore we must work close to it. We denote by
$v_{\varphi }(\omega )$ and $v_{\varphi }(2\omega )$ the phase
velocities of the fundamental and second harmonic respectively.
The phase-matching condition obviously~is
\begin{equation}
v_{\varphi }(\omega )=v_{\varphi }(2\omega ).
\end{equation}
It is well-known \cite{yariv} that this phase-matching condition
can be realized by some adequate choice of the value of the angle
$\theta $ between the optical axis ($z$-axis) and the propagation
direction. Several situations arise, depending on the relative
magnitude of the indices in the considered material and on the
polarization of both waves. We consider only one of these cases,
of major importance for applications. For KDP and many other
materials, the inequalities $n_{o}(2\omega )>n_{o}(\omega )$ and
$n_{o}(\omega )>n_{e}(\omega )$ are satisfied. It is easily shown
that in this case the phase-matching angle exists when the
fundamental wave is ordinary and the second-harmonic
extraordinary, but does not exist for other polarizations, except
if both waves are extraordinary. In the latter situation, the
experimental values of $n_{e}$ and $n_{o}$ do not yield a real
phase-matching angle value. We restrict thus the study to the
particular case where the fundamental wave is ordinary and the
second-harmonic extraordinary, which is the only physical
situation in which the phase-matching angle exists for the materials under
consideration. The phase-matching condition is then  \cite{yariv}
\begin{equation}
\sin ^{2}\theta =\frac{n_{o}^{-2}(\omega )-n_{o}^{-2}(2\omega )}
{n_{e}^{-2}(2\omega )-n_{o}^{-2}(2\omega )}.  \label{2}
\end{equation}
The first condition to be satisfied by our ansatz is that the angle $\theta $
between the propagation direction and the $z$-axis must satisfy the
phase-matching condition~(\ref{2}), at least in an approximate way.

\subsection{Transverse velocity}

The dispersion relation $\omega (\vec{k})$ follows from (\ref{43}), the
value of the group velocity is then computed according to
 $\vec{v}={\partial \omega }/{\partial \vec{k}}$. While the group velocity of the
ordinary wave is parallel to the phase velocity, that of the extraordinary
wave has a transverse component~\cite{yariv}. It is
\begin{equation}
v_{e,x}=\frac{c^{2}k}{2\Lambda }\left( \frac{1}{n_{e}^{2}}-\frac{1}{n_{o}^{2}}\right) \sin 2\theta,
\end{equation}
with
\begin{equation}
\Lambda =\omega ^{2}+c^{2}k^{2}\left( \frac{\cos ^{2}\theta \;n_{o}^{\prime }}
{n_{o}^{3}}+\frac{\sin ^{2}\theta \;n_{e}^{\prime }}{n_{e}^{3}}\right)
\end{equation}
($k$ is the wave vector and $\omega $ the pulsation of the fundamental,
$n_{o}^{\prime }={dn_{o}}/{d\omega }$). Because we expect that the
walk-off is small, $v_{e,x}$ must be close to zero. This can be achieved in
two ways: either by setting $n_{e}$ close to $n_{o}$, or when $\sin 2\theta $
is small. Try first to satisfy the former condition: anisotropy must
be weak. However, anisotropy is needed for the phase-matching that must be
realized at least in an approximate way. The difference $(n_{e}-n_{o})$
appears indeed in the denominator of (\ref{2}), and therefore it cannot be
zero. The anisotropy, or the difference $(n_{e}-n_{o})$, must be small but not
too small. This is the second condition imposed to our ansatz. It can be
satisfied in many real materials.

It is likely also necessary that $\sin2\theta$ is small. We note in
(\ref{2}) that, when $n_e(2\omega)=n_o(\omega)$, then $\sin^2\theta=1$. In this
case $\theta =\pi/2$ and $\sin 2\theta=0$. We deduce this way a~third
condition to be satisfied by the ansatz: $n_e(2\omega)$ must be close to
$n_o(\omega)$.
Thus the achievement of phase-matching and reduction of walk-off necessitate
that $n_o(2\omega)$ and $n_e(\omega)$ are very close together, and that
anisotropy is weak. The ansatz used below is built in order to satisfy these
 conditions.  It is seen
from published experimental data~\cite{dmitriev} that they are
realized in many real materials which are in fact the most commonly  used
 for practical applications of the second harmonic generation.
Further it was very early recognized that
the arising of the transverse group velocity or walk-off is the main limitation to
an efficient second harmonic generation at phase-matching angle \cite{boy65a}, and
the fact that the condition ``$n_e(2\omega)$ close to
$n_o(\omega)$'' ensures the reduction of the walk-off in this situation follows straightforwardly
from the properties of light
 propagation in a birefringent medium.

\section{An adequate ansatz}

\subsection{The scaling parameter}

As usual in the frame of multiscale analysis, we introduce some small
parameter $\varepsilon $. It characterizes the pulse size in the
following way: we call $L$ the common order of magnitude of the pulse
length and width, and $\lambda $ the wavelength. Then the perturbative
parameter is defined as
\begin{equation}
\varepsilon =\frac{\lambda }{L} \,. \label{epsi}
\end{equation}
The electric field $\vec{E}$ is expanded both in a power series of some
small parameter~$\varepsilon $, and in a Fourier series of some fundamental
phase~$\phi $ as
\begin{equation}
\vec{E}=\sum_{l\geq l_{0},p\in {\mathbb Z}}\varepsilon ^{l}
\vec{E}_{l}^{p}e^{ip\phi } . \label{3}
\end{equation}
$l_{0}$ fixes the order of magnitude of the main term.  It is discussed
below. The amplitudes $\vec{E}_{l}^{p}$ are functions of slow variables that
are defined below. The main slow variables characterize the pulse
shape and its propagation at the group velocity. According to (\ref{epsi}),
they are variables of order $\varepsilon $, defined by
\begin{equation}
\xi =\varepsilon x, \qquad\eta =\varepsilon y,\qquad
\zeta =\varepsilon z,\qquad  \tau =\varepsilon t.
\label{4.2a}
\end{equation}
We seek for the evolution of the pulse shape through propagation distances
long in relation to the pulse length $L$, say $L/\varepsilon $, as in the
standard nonlinear Schr\"{o}dinger model.
On these grounds and  because the propagation direction is close to the $x$-axis (see below),
the slow variable giving account
for the evolution can be chosen as
\begin{equation}
\xi _{2}=\varepsilon ^{2}x  .\label{4.2b}
\end{equation}

\subsection{The carrier}

The phase $\phi $ determines the carrier monochromatic plane wave. It thus
determines the propagation direction. Recall that, because phase-matching
must be at least approximately realized, the angle $\theta $ between the
 $z$-axis and the propagation direction must be close to the value given in
(\ref{2}). On the other hand, it has been shown in subsection 2.3 that the latter
value must be close to $\pi /2$. We write it as
\begin{equation}
\theta =\frac{\pi }{2}-\varepsilon ^{a}\gamma , \label{tet}
\end{equation}
where $\gamma $ is a free parameter at this point, and the exponent $a$ is
positive and is determined below. The phase $\phi $ is
\begin{equation}
\phi =\vec{k}\cdot \vec{x}-\omega t , \label{phi0}
\end{equation}
with $\vec{k}=k(\sin \theta ,0,\cos \theta )$. Inserting the definition
(\ref{tet}) of $\theta$ into the expression (\ref{phi0}) of $\phi$, we would get
\begin{equation}
\phi =kx+k\gamma \varepsilon ^{a}z-\omega t+O\left(\varepsilon ^{2a}\right).
\label{phi0bis}
\end{equation}
A dependency with regard to some slow variable $\varepsilon ^{a}z$
appears ($a>0$). For $a\geq 1$, this dependency can be
incorporated in the amplitude, that depends on $\zeta =\varepsilon
z$. Thus only values of $a$ less than 1 are interesting. On these
grounds, and taking into account the fact that the perturbative
expansion will involve only integer powers of $\varepsilon^a$, it
is seen that the correct choice for $a$ is  $a=1/2$. Further, due
to the anisotropy, the wave vector norm $k$ depends on the angle
$\theta$. Thus, according to (\ref{tet}),
 $k$ depends on $\varepsilon $. As $\varepsilon $ tends to 0, $k$
must be expanded in a power series of $\sqrt{\varepsilon }$. A term
involving $\sqrt{\varepsilon }x$ appears then in (\ref{phi0bis}). The terms
of order $\varepsilon ^{1}$ and higher are incorporated in the amplitude.
The complete phase $\phi $ can thus be written  as
\begin{equation}
\phi =k\left( 1+\alpha \sqrt{\varepsilon }\right) x+\sqrt{\varepsilon }k\gamma z-\omega t.  \label{4}
\end{equation}
In the same way as $k$, all quantities such as group velocities or
polarization vectors, that depend on the angle $\theta$, are  expanded in
a power series of $\sqrt\varepsilon$. Therefore we must introduce an
additional slow variable of half-integer order,
\begin{equation}
\xi _{1}=\varepsilon ^{\frac{3}{2}}x . \label{4.2c}
\end{equation}
The complete set of slow variables on which the amplitudes $\vec{E}_{l}^{p}$
depend is thus\linebreak $(\xi,\eta,\zeta,\tau,\xi_1,\xi_2)$. They are
defined by (\ref{4.2a}), (\ref{4.2b}), and (\ref{4.2c}).

Leaving aside the half integer powers of $\varepsilon $, this is
mainly the ``classical'' multiscale expansion that leads ordinary
to the nonlinear Schr\"{o}dinger-type (NLS-type) models
(\cite{dodd}, p.~495 {\it sq.}) describing long-distance
propagation along the $x$-axis. Recall indeed that it is not
necessary to introduce {\it a priori} the statement that the field
is at first order a function of $(x-Vt)$. Else, the linear
transport equation, that is the equation giving account for the
modulation propagation at group velocity, is obtained as a
solvability condition of the perturbative scheme. The main
originality of the expansion (\ref{3})--(\ref{4.2c}) is the
introduction of an intermediate scale $x\sqrt{\varepsilon}$,
$z\sqrt{\varepsilon}$ in the expression of the phase $\phi$. The
expansion therefore involves not only 3 scales for the
longitudinal space variable, as in the usual NLS-type expansions,
but 5 different scales (including $\xi_1$). The term
$\sqrt\varepsilon k\gamma z$ in expression~(\ref{4}) for $\phi$
represents a deviation for an angle $\sqrt\varepsilon\gamma$
 from the propagation along the $x$-axis.

\subsection{Expansion of the susceptibility}

Here comes the difficult point which is the introduction of a second approximation beside
the multiscale expansion. It has been seen that the components of the linear
susceptibility tensor must satisfy several hypotheses that allow certain
approximations to be made.
The problem is that the only consistent way of describing
mathematically these hypotheses requires the $\chi ^{(1)}$
tensor to be written as a function of the perturbative parameter $\varepsilon
$. However,
 $\varepsilon $ represents the pulse size, and $\chi ^{(1)}$ does not depend
on it from the physical point of view. Nevertheless, we assume that the
linear susceptibility tensor $\chi ^{(1)}$ can be formally expanded in a
power series of $\sqrt{\varepsilon }$ in the following way:
\begin{equation}
1+\hat{\chi}^{(1)}=n^{2}+\sqrt{\varepsilon }\hat{\chi}_{\frac{1}{2}}^{(1)}+\varepsilon \hat{\chi}_{1}^{(1)}+\varepsilon ^{\frac{3}{2}}
\hat{\chi}_{\frac{3}{2}}^{(1)}+\cdots . \label{1}
\end{equation}
$n^{2}$ is an isotropic index, and the fact that
$\lim\limits_{\varepsilon \rightarrow 0}\left(
1+\hat{\chi}^{(1)}\right) =n^{2}$ gives a partial account for the
second condition of \S~2.3, the weak anisotropy hypothesis. A
particular definition of the coefficients $\hat{\chi}_{j}^{(1)}$
can be given. It is of no interest for our purpose, and is left
for further publication~\cite{match2}.

\subsection{The amplitude}

An important difference between the ansatz (\ref{3}) and the standard one
for NLS is the order of magnitude of the leading term, that is the value of
$l_{0}$ in (\ref{3}).
It is determined as follows. The model that we intend to derive and to justify
is used in \cite{torner,menyuk,kanashov}. It has the
following form:
\begin{subequations} \label{5}
\begin{gather}
i\partial _{\xi _{2}}\varphi +D_{\vec{\xi}}^{2}\,\varphi   =  A_{1}\,\psi
\varphi ^{*},\\
i\partial _{\xi _{2}}\psi +{D^{\prime }}_{\vec{\xi}}^{2}\,\psi   =
A_{2}\,\varphi ^{2},
\end{gather}
\end{subequations}
where ${D}_{\vec{\xi}}^{2}$ and ${D^{\prime }}_{\vec{\xi}}^{2}$
are second-order partial differential operators relative to the
slow variab\-les
 $\left(
\tau -{\xi }/{V}\right)$, $\eta $, $\zeta $ that describe the
shape of the pulse.  $A_{1}$, $A_{2}$ are some constants. $\varphi
$~and $\psi $ are the amplitudes of the fundamental and second
harmonic respectively. These amplitudes are assumed to have an
order of magnitude $\varepsilon ^{l_{0}}$. Furthermore, $\xi $,
$\eta $, $\zeta $, $\tau $ have an order of magnitude $\varepsilon
$, and $\xi _{2}$ an order $\varepsilon ^{2}$. Therefore the
left-hand-side of equation~(\ref{5}) has the order $\varepsilon
^{2+l_{0}}$, and its right-hand-side has the order $\varepsilon
^{2l_{0}}$. We must thus have $2+l_{0}=2l_{0}$, and $l_{0}=2$.

This differs from the case of the cubic NLS equation, which is
\begin{equation}
i\partial _{\xi _{2}}f+D_{\vec{\xi}}^{2}f+Af|f|^{2}=0, \label{6}
\end{equation}
 with
analogous notations, $f$ being the modulated amplitude.
There we would have $l_{0}+2=3l_{0}$, thus $l_{0}=1$. This has a very
important physical meaning: the present model requires input energy pulses
from a smaller range than the NLS solitons obtained either in cubic media by
the Kerr effect or in quadratic media by cascading, far from phase-matching~\cite{kal94b}.
This is due to the fact that the interaction is resonant, or very close to
resonance. Note that the multiscale analysis justifies this difference
between the required power inputs without making any
special assumption about the values of the nonlinear susceptibilities.

\section{Solving the perturbative scheme}

\subsection{A technical point}

The order by order resolution of the perturbative scheme can now be
performed. The detail of this derivation is given in the appendix. The
vector amplitudes whose evolution is under investigation are polarized so that
\begin{equation}
\vec{E}_{2}^{1}=\left(
\begin{array}{c}
0 \\
\varphi  \\
0
\end{array}
\right), \qquad \vec{E}_{2}^{2}=\left(
\begin{array}{c}
0 \\
0 \\
\psi
\end{array}
\right).   \label{10}
\end{equation}
The transport equation giving the group velocity is expected to appear at
order $\varepsilon ^{l_{0}+1}=\varepsilon ^{3}$. We find that the amplitude $\varphi $ of the
fundamental satisfies the partial differential equation
\begin{equation}
\partial _{\xi }\varphi +\frac{1}{v}\partial _{\tau }\varphi =i\delta
_{1}\varphi.   \label{12}
\end{equation}
The constants $\delta _{1}$ and $v$ are specified below (equation
(\ref{18})). The solution of equation~(\ref{12}) is
\begin{equation}
\varphi (\xi ,\xi _{1},\xi _{2},\tau ,\eta ,\zeta )=\varphi _{0}\left( \xi
_{1},\xi _{2},\tau -\frac{\xi }{v},\eta ,\zeta \right) \,e^{\displaystyle
i\delta _{1}\xi }.  \label{13}
\end{equation}
It describes both a propagation at speed $v$ and a linear phase modulation
measured by the coefficient $\delta _{1}$.

The expressions of the constants $v$ and $\delta _{1}$ are given by solving
the multiscale expansion. They can be interpreted, and written in a
convenient way, by using some power series expansion of the wave numbers.
We introduce the following notations:
\begin{gather}
\tilde{n}_{s}^{2}=n^{2}+\sqrt{\varepsilon }\Delta n_{s}^{2}+
\varepsilon \hat{\chi}_{1,s}^{(1)}+
\varepsilon ^{\frac{3}{2}}\hat{\chi}_{\frac{3}{2},s}^{(1)}+\cdots \qquad
\mbox{for  $s=o$ or $e$},  \label{14}
\\
\tilde{k}_{o}=\tilde{k}_{o}(\omega )=\frac{\omega \tilde{n}_{o}(\omega )}{c}
=k_{0,o}+\sqrt{\varepsilon }k_{\frac{1}{2},o}+\varepsilon k_{1,o}+\cdots,
\label{15}
\\
\tilde{k}_{e}=\tilde{k}_{e}(2\omega )=\frac{2\omega \tilde{n}_{e}(2\omega )}{c}
=k_{0,e}+\sqrt{\varepsilon }k_{\frac{1}{2},e}+\varepsilon k_{1,e}+\cdots.
\label{16}
\end{gather}
For the derivatives of these quantities, the  notations used are
\begin{equation}
k_{p,o}^{\prime }=\frac{d}{d\omega }k_{p,o}(\omega )\qquad
\mbox{and}
\qquad
k_{p,e}^{\prime }=\frac{d}{d(2\omega )}k_{p,e}(2\omega ).
\label{17}
\end{equation}
Then the values of the coefficients are found,
\begin{equation}
\delta _{1}=k_{1,o}-\frac{k_{0,o}\gamma ^{2}}{2}\hspace{1cm}\mbox{ and }
\hspace{1cm}\frac{1}{v}=k_{0,o}^{\prime } . \label{18}
\end{equation}
Thus the phase factor $e^{i\delta _{1}\xi }$  accounts  for the
corrections to the $x$-component of the wave vector $\vec{k}$ at
order $\varepsilon $, in norm (term $k_{1,o}$) and in direction
(${\rm 2^{nd}}$ term). Similar corrections appear at following
order $\varepsilon ^{3+\frac{1}{2}}$, it is the factor $e^{i\rho
_{1}\xi _{1}}$ in formula (\ref{27}), and for the second harmonic:
the factors $e^{i\rho _{1}\xi _{1}}$ and $e^{i\rho _{2}\xi _{1}}$
in formulas (\ref{27b}) and (\ref{30}). These corrections are a
characteristic feature of the kind of ``mixed'' expansion used
here.

Note that it is decisive, for going further in the perturbative expansion,
that equation~(\ref{12}) has its particular form. The  multiscale
expansion frame considered describes indeed the long-distance evolution of some state,
that is a steady state in the first approximation. Usually, the order zero
is a plane wave, invariant in amplitude, thus order one can be considered.
The order one presents Galilean invariance, thus order two can be
considered. Higher order make sense mainly if order two describes a soliton,
which is another kind of invariance. Therefore, the arising of a $\xi _{1}$-dependency
 should {\it a priori} forbid to go to the next order, except if
this dependency presents some invariance. As an example, an exponentially
decreasing $\xi _{1}$-dependency, describing absorption, would reduce all
following orders to small linear corrections to this decreasing, that would
give account for the main physical behavior. There is, however, some invariance
here, because the $\xi _{1}$-dependent factors are unitary phase factors,
and we can pursue the expansion.

\subsection{The phase-matching condition}

The first order of the dispersion relation, found at first nonzero order $\varepsilon^2$, is
\begin{equation}
k=\frac{n(\omega)\omega}c=\frac{n(2\omega)\omega}c . \label{7}
\end{equation}
Because $k$ is unique, the first order of the phase-matching condition is
obtained as
\begin{equation}
n(2\omega)=n(\omega) . \label{8}
\end{equation}
The common value of $n(\omega)$ and $n(2\omega)$ is written $n$ below.

At order $\varepsilon ^{2+\frac{1}{2}}$, the anisotropy appears through a
condition involving $\alpha k$, the term of order $\sqrt{\varepsilon }$ in
the expansion of $k$. We choose, as written above, the fundamental and the
second harmonic as ordinary and extraordinary waves respectively.  $\alpha $ takes the value
\begin{equation}
\alpha =\frac{\Delta n_{o}^{2}(\omega )}{2n^{2}}=\frac{\Delta
n_{e}^{2}(2\omega )}{2n^{2}} . \label{9}
\end{equation}
Because $\alpha $ is unique,
the second order of the phase-matching condition is got at this point. It
is
\begin{equation}
\Delta n_{o}^{2}(\omega )=\Delta n_{e}^{2}(2\omega ).  \label{11}
\end{equation}

Thus the two first orders of the phase-matching condition are a direct
consequence of the ansatz, precisely of the unicity of the wave vector in
the phase $\phi$ defined by~(\ref{4}). A~more complicated ansatz involving
two fundamental phases could be envisaged. However, it would not be able to give
account for the wave interaction. It is indeed proved below that the
following terms in the expansion of the linear susceptibility $\hat{\chi}^{(1)}$
 must also coincide (see equation (\ref{44})). The same kind of
considerations would prove the same result for first order, if it was not
set {\it a priori} in the ansatz.

The following order of the phase-matching condition appears at order $\varepsilon ^{4}$
 when wri\-ting down the nonlinear evolution equation for the
complex amplitudes $\varphi $ and $\psi $ of the fundamental and second
harmonic respectively. They are equations (\ref{5}), while the
amplitudes~$\varphi $ and~$\psi $ satisfy
\begin{subequations}
\begin{gather}
\varphi =\varphi _{1}\left( \xi _{2},\tau -\frac{\xi }{v}-
\frac{\xi _{1}}{v_{1}},\eta ,\zeta -\gamma \xi _{1}\right) \,e^{i(\delta _{1}\xi +\rho
_{1}\xi _{1})} , \label{27b}
\\
\psi =\psi _{1}\left( \xi _{2},\tau -\frac{\xi }{v}-\frac{\xi _{1}}{v_{1}},\eta ,\zeta -\gamma \xi _{1}\right)
 \,e^{i(\delta _{2}\xi +\rho _{2}\xi_{1})}  .\label{30}
\end{gather}
\end{subequations}
Equations (\ref{5}) are expressed in a coordinate frame that moves with the
wave, at a speed equal to the group velocity of the wave up to order $\sqrt{\varepsilon }$.
 The coordinates are defined by
\begin{equation}
 \tilde{\tau}=  \tau -\frac{\xi }{v}-\frac{\xi_{1}}{v_{1}} \,,\qquad
\tilde{\eta}= \eta, \qquad
\tilde{\zeta}=\zeta -\gamma \xi _{1},\qquad\tilde{\xi}_{2}= \xi _{2}.
 \label{elene}
\end{equation}
In this frame, there is no dependency of the amplitudes relative
to the variables $\xi $ and $\xi _{1}$.
 On the other hand, phase factors depending on these variables
appear. These factors are
\begin{equation}
\exp {i\left[ (\delta _{2}-2\delta _{1})\xi +(\rho _{2}-2\rho _{1})\xi
_{1}\right] }
\end{equation}
in the evolution equation for $\varphi $, and the inverse in the evolution
equation for $\psi $. These two equations yield equations (\ref{33b}) and (\ref{41}) respectively,
 after cancellation of the factors mentioned. This
yields two approximate phase-matching conditions, in addition to (\ref{9}):
 $\delta _{2}=2\delta _{1}$ and $\rho _{2}=2\rho _{1}$. The first condition can be written as
$k_{1,e}=2k_{1,o}$, or as
\begin{equation}
\chi _{e}^{(1)}(2\omega )=\chi _{o}^{(1)}(\omega ) . \label{44}
\end{equation}

\subsection{The phase-matching angle}

Conditions (\ref{8}), (\ref{11}) and (\ref{44}) do not involve a
phase-matching angle. They state that the approximate phase-matching must be
realized ``spontaneously'' by the material up to order $\varepsilon $. The
last of the phase-matching conditions obtained in the previous subsection~is
\begin{equation}
\rho_2=2\rho_1.
\end{equation}
It can be satisfied by some particular choice of the angle $\sqrt\varepsilon\gamma $ according to
\begin{equation}
\gamma^2=\frac{nc}\omega\frac{k_{\frac32,e}(2\omega)-2k_{\frac32,o}(\omega)}
 {\Delta n_o^{2}(\omega) -\Delta n_o^{2}(2\omega) }\, . \label{45}
\end{equation}
Formula (\ref{45}) coincides with the usual phase-matching condition (\ref{2}),
 taking into account the power series expansion of $\chi^{(2)}$ and the
``spontaneous'' approximate phase-matching, up to order $\varepsilon$. A deviation
from the exact phase-matching of order $\varepsilon^2$ is possible. It is
described by the terms $B_1\varphi$ in equation (\ref{33b}) and $B_2\psi$ in equation (\ref{41}).
 The present multiscale expansion proves that the model is by no
means valid for a larger deviation. In such a case, only a model involving
non-resonant interaction can describe the physical phenomenon. Completely
integrable models such as the nonlinear Schr\"odinger equation (NLS) or the
Davey Stewartson system (DS I and II) can be derived in this frame, but
involve much higher intensities: of order $\varepsilon$ instead of $\varepsilon^2$
 in the present case~\cite{dsopt1,dsopt2}.

\subsection{The group velocity matching}

The phase-matching is a phase velocity matching, but for short
pulses a group velo\-ci\-ty matching is also necessary. Without
it, the two pulses cross each other without ha\-ving time enough
to interact. Mathematically, the group velocity matching condition
appears through the requirement that the variables
$(\tilde\tau,\tilde\eta,\tilde\zeta,\tilde\xi_2)$ are uniquely
defined by equation~(\ref{elene}). In other words, the frame
moving with the pulses, in which equation~(\ref{5}) describes the
evolution of the waves, must be defined in a unique way, thus must
be the same for both waves.

At order $\varepsilon^3$, the first order term of the group velocity is
found. It is $1/v=k^{\prime}_{0,o}$ for the fundamental and
$1/v=k^{\prime}_{0,e}$ for the second harmonic. According to the above
statements, the velocity $v$ is the same for both waves.
If these velocities differ, it is found at next order that no interaction
between the waves is possible. From the formal point of view, the $\xi_1$
and $\xi_2$-dependencies of $\varphi$ and $\psi$ would be linear if both
group velocities have not the same value. For this reason, we must have
$k_{0,e}^{\prime}= k_{0,o}^{\prime}$: this means that the group velocities
coincide at order zero. Using the zero order phase-matching condition (\ref{8}),
 this condition can be written in terms of the index $n$ as
\begin{equation}
n^{\prime}(\omega)=2 n^{\prime}(2\omega).  \label{21}
\end{equation}

The first correction $1/{v_{1}}$ to the inverse of the group velocity
is obtained at following order $\varepsilon ^{3+\frac12}$. As at previous
order, this correction is ${1}/{v_{1}}=k_{\frac{1}{2},o}^{\prime }$ for the
fundamental and ${1}/{v_{1}}=k_{\frac{1}{2},e}^{\prime }$ for the second
harmonic. Therefore $k_{\frac{1}{2},o}^{\prime }=k_{\frac{1}{2},e}^{\prime }$
or, in terms of the indices,
\begin{equation}
\left(\Delta n_{o}^2\right)^{\prime }(\omega )=2\left(\Delta n_{e}^2\right)^{\prime }(2\omega ) . \label{32}
\end{equation}
An explicit walk-off appears in the final equations (\ref{33b})--(\ref{41})
through the terms involving a first order $\tau $-derivative. It is measured
by the difference between~$k_{1,o}^{\prime }$ and~$k_{1,e}^{\prime }$. As
the deviation from phase-matching, the walk-off has a fixed maximal order of
magnitude, that of $\varepsilon $. If this group velocity matching is not
satisfied, the model yielded by equations (\ref{33b})--(\ref{41}) is
theoretically no more valid. This means physically that for a pulse size,
a power input, and a propagation distance in accordance with the scaling,
the two waves do not coexist at the same place during a long enough time,
and cannot interact.

\subsection{The asymptotic model}

As mentioned above, the evolution equations of the form (\ref{5}) for the
complex amplitu\-des~$\varphi $ and~$\psi$ of the fundamental and second
harmonic respectively are obtained at order~$\varepsilon^4$. They reduce to
\begin{subequations}
\begin{gather}
\biggl[ 2ik\partial_{\xi_2}-B_1+2ik\left(k_{1,o}^{\prime}+
\frac{\gamma^2}{2v}\right) \partial_\tau \nonumber\\
\phantom{\biggl[ 2ik\partial_{\xi_2}}{} -kk^{\prime\prime}\partial_\tau^2+\partial_\eta^2
+\partial_\zeta^2-2ik\alpha\gamma\partial_\zeta\biggl]  \varphi =A_1\,
\psi\varphi^\ast,
\label{33b}\\
\biggl[ 4ik\partial _{\xi _{2}}-B_{2}+4ik\left( k_{1,e}^{\prime
}+\frac{\gamma ^{2}}{2v}\right) \partial _{\tau }\nonumber\\
\phantom{\biggl[ 4ik\partial _{\xi _{2}}}{} -\frac{2ik\gamma }{n^{2}}\left( 2\Delta n_{o}^{2}+\Delta
n_{e}^{2}\right) \partial _{\zeta }-2kk_{0,e}^{\prime \prime }\partial
_{\tau }^{2}+\partial _{\eta }^{2}+\partial _{\zeta }^{2}\biggr] \psi
=A_{2}\,\varphi ^{2}.
\label{41}
\end{gather}
\end{subequations}
(The notations $\Delta n_{s}^{2}$, $k_{j,s}$ ($s=o,\,e$) are defined by
equations (\ref{14}) to (\ref{16})).
 The coefficients, $2ik$ or $4ik$ of $\partial _{\xi _{2}}$, and
$-kk^{\prime \prime }$ ($=-k_{0,o}k_{0,o}^{\prime \prime }$) or
$-2kk_{0,e}^{\prime \prime }$ of $\partial _{\tau }^{2}$, are the
usual coefficients in the NLS models in nonlinear optics. A
transverse velocity term, proportional to $\partial _{\zeta
}\varphi $, and a longitudinal one, proportional to $\partial
_{\tau }\varphi $, appear (the coefficient $\alpha $ is given by
equation~(\ref{9})). $k_{1,o}^{\prime }$ and $k_{1,e}^{\prime }$
are the corrections to the inverse of the group velocities at
order~$\varepsilon $.

The coefficients $B_{1}$ and $B_{2}$ are some corrections to the phase
vector at this order. They have the expressions
\begin{gather}
B_{1}=-2kk_{2,o}+\frac{\omega ^{2}}{2c^{2}}\left[ \frac{3\left(\Delta
n_{o}^2\right)^2(\omega )}{4n^{4}}-\hat{\chi}_{1,o}^{(1)}(\omega )\right] \gamma
^{2}+\frac{k^{2}}{4}\gamma ^{4} , \label{34b}
\\
 B_{2}=-4kk_{2,e}+4k^{2}\biggl[ \frac{1}{2n^{2}}
\left( \hat{\chi}_{1,e}^{(1)}-2\hat{\chi}_{1,o}^{(1)}\right) \nonumber\\
\qquad\qquad \qquad\qquad\quad{}-\frac{1}{8n^{4}}\left( \left(\Delta n_{e}^2\right)^2+4\Delta
n_{o}^{2}\Delta n_{e}^{2}-8\left(\Delta n_{o}^2\right)^2\right) \biggr] \gamma^{2}+k^{2}\gamma ^{4}.
\label{42}
\end{gather}
The nonlinear interaction constant $A_{1}$ and $A_{2}$ are defined
in the general case by equations (\ref{35b})--(\ref{35t}) (in
appendix). Using the symmetry properties of the $\chi
^{(2)}$-tensor, inclu\-ding the so-called complete symmetry
property, we have $A_{2}=2A_{1}$, at least for the $\bar{4}2{\rm
m}$
 and 3m symmetry classes, for which the common value of these
coefficients is given by equations~(\ref{38}) and~(\ref{38l}).

Equations (\ref{33b}) and (\ref{41}) yield the sought model. They are
analogous to the equations given by Menyuk in \cite{menyuk}, but here in a
complete $(3+1)$-dimensional version. Kanashov and Rubenchik shown~\cite{kanashov}
that no collapse occurs in this model, and that a soliton
solution exists in the sense of a stable localized pulse. System (\ref{33b})--(\ref{41})
 is written  in a coordinate frame that moves with the wave at a
speed equal to its group velocity up to order $\sqrt{\varepsilon }$.  It is
seen that the group velocity of both fundamental and second harmonic must
coincide up to this order. Walk-off terms of order $\varepsilon $ appear in
equations (\ref{33b}) and~(\ref{41}).  These are the terms involving first
order $\tau $- and $\zeta $-derivatives. In an analogous way, it is seen
that the approximate phase-matching must be realized ``spontaneously'' by
the material up to order $\varepsilon $, and once more at order $\varepsilon
^{3/2}$ by some particular choice of the angle $\sqrt{\varepsilon }\gamma $,
according to equation~(\ref{45}).

Consider now the expansion (\ref{1}) of the susceptibility tensor, that
also has the expression (\ref{43}). Because we must have an isotropic zero
order term, $\sqrt{\varepsilon }$ is the order of magnitude of the
anisotropy. The interaction must be phase-matched ``spontaneously'' up to
order $\varepsilon $, thus the order of magnitude of the difference
$\left(n_{o}(\omega )-n_{e}(2\omega )\right)$ must be~$\varepsilon ^{3/2}$. Last, the
group velocity of both waves must be identical up to order $\sqrt{\varepsilon
}$.
 Thus their difference must have an order of magnitude $\varepsilon $.
Finally, the matching conditions are
\begin{subequations}
\begin{gather}
 n_{o}^{2}(\omega )-n_{e}^{2}(\omega )  \in  O\left(\sqrt{\varepsilon }\right), \\
 \frac{d\,n_{o}^{2}(\omega )}{d\omega }-\frac{d\,n_{e}^{2}(2\omega )}{d\omega }  \in O(\varepsilon ),
\\
 n_{o}^{2}(\omega )-n_{e}^{2}(2\omega )  \in O\left(\varepsilon^{3/2}\right).
\end{gather}
 \label{46}
\end{subequations}
$\omega _{0}$ is the fundamental frequency and $\varepsilon $ the
perturbation parameter.

\section{Conclusion}

A long-distance propagation model (equations (\ref{33b})--(\ref{41})) for two
interacting waves with different velocities has been derived through a
rigorous multiscale analysis. It uses a new  type of expansion that mixes a
multiscale expansion, and an expansion in a power series of some physical
quantity, that here represents mainly the anisotropy. This expansion
involves 5 longitudinal space
scales, instead of 3 in the usual NLS-type models. The supplementary scale
is related to approximate conditions satisfied by the linear
susceptibilities. The expansion considered will be useful for the
mathematical justification of asymptotic models of the above mentioned type.
This rigorous derivation has allowed the determination of the validity
conditions of the model.

From a more concrete point of view, the model equations
(\ref{33b}-\ref{41}) describe the evolution of the modulation of
two short localized pulses, fundamental and second harmonic,
propagating together in a bulk uniaxial crystal with non-vanishing
second order susceptibility $\chi^{(2)}$, and interacting. Much
theoretical work has already been done about this system. It has
been proved that no collapse occurs \cite{kanashov,berge},
analytical solutions have been found in $(1+1)$ dimensions~\cite{menyuk},
and numerical simulations have shown the existence
of stable solutions~\cite{torner}.
 We have determined the physical conditions under which the model
is valid: both phase and group velocity of the waves must be close enough
together. Group velocities are matched when anisotropy is small.  Thus the
phase-matching angle must be small. Therefore, the phase-matching must be
realized by the intrinsic properties of the material and for any propagation
direction with some precision. Hence the model cannot be valid in any
material and at any frequency.
X-Mozilla-Status: 0000

\appendix

\section*{Appendix. Details of the derivation}

\setcounter{section}{1}

\subsection{Dispersion relation}

In this appendix, we give the rigorous derivation of system (\ref{33b})--(\ref{41}),
using the perturbative scheme described in Section~2. The
convolutions involved by expression (\ref{de}) of $\vec{D}$ are treated
using the formulas derived in~\cite{dsopt1}. We solve it order by order:
the first nonzero order, $\varepsilon ^{2}$, gives that $E_{2}^{p,s}$ is zero for
all $s=x,y,z$, and all integer $p$, except for $p=\pm 1,\,\pm 2$, and
$s=y,z$, if the dispersion relation (\ref{7}) is satisfied.

At order $\varepsilon^{2+\frac12}$, a first correction to the dispersion
relation is obtained through the expression (\ref{9}) of the coefficient $\alpha$.
 The polarization vectors $\vec E_2^1$ and $\vec E_2^2$ are also
found, they are defined as (\ref{10}). $\varphi$ and $\psi$ are the
amplitudes whose evolution is under investigation, and the other $\vec E_2^p$
are zero (except for the symmetric terms $p=-1,-2$).

\subsection{Order 3 and group velocity}

At order $\varepsilon^3$, the group velocity appears. Precisely, we get as a
solvability condition for $\varphi$ the equation (\ref{12}), which admits
the solution (\ref{13}). In an analogous way, the $\xi$-dependency of the
second harmonic is obtained:
\begin{equation}
\psi=\psi_0\left(\xi_1,\xi_2,\tau-\frac\xi v,\eta,\zeta\right) \, e^{ i\delta_2\xi}  ,\label{19}
\end{equation}
with
\begin{equation}
\delta_2=k_{1,e}-\frac{k_{0,e}\gamma^2}2 \qquad \mbox{and} \qquad
\frac1v=k_{0,o}^{\prime}.  \label{20b}
\end{equation}

Other conditions give
\begin{equation}
\vec E_{2+\frac12}^1=\left(
\begin{array}{c}
0 \\
f \\
0
\end{array}
\right),\qquad \vec E_{2+\frac12}^2=\left(
\begin{array}{c}
-\gamma\psi \\
0 \\
g
\end{array}
\right) . \label{22}
\end{equation}
The functions $f$ and $g$ have to be determined.

\subsection{Order $\boldsymbol{\varepsilon^{3+\frac12}}$}

The following order $\varepsilon^{3+\frac12}$ gives, in a similar way, the
term of order $\sqrt\varepsilon$ in the group velocity. The following
equation is then obtained:
\begin{gather}
-2ik\partial_\xi f
-2ik\alpha\partial_\xi\varphi-2ik\partial_{\xi_1}\varphi=
2k\delta_1 f+\frac{2ik}V \partial_\tau f \nonumber\\
\phantom{-2ik\partial_\xi f}+\frac{\omega^2}{c^2}\hat\chi_{\frac32,o}^{(1)} \varphi+
\frac{i\omega}{c^2}\left( \left(\Delta
n_o^2\right)^{\prime}\omega+ 2\Delta n_o^2\right)\partial_\tau\varphi.
\label{23}
\end{gather}
Equation (\ref{23}) is solved as follows: using the change of variables
\begin{equation}
\hat \tau =\tau-\frac\xi v ,\qquad
\hat\xi = \xi,
  \label{24}
\end{equation}
noting that $\varphi=\varphi_0(\hat\tau)e^{i\delta_1\hat\xi}$ (equation (\ref{13})),
and setting
\begin{equation}
f=f_0(\hat\xi,\hat\tau)e^{i\delta_1\hat\xi}  \label{25}
\end{equation}
in equation (\ref{23}), it is found that $\partial_{\hat\xi}f_0$ expresses in
terms of $\varphi_0$ and of its derivatives. Thus, relative to $\hat\xi$,
$\partial_{\hat\xi}f$ is some constant $A$, and
\begin{equation}
f_0=A\hat\xi+B,
\end{equation}
$B$ being some constant. Because $f_0$ must be bounded as $\hat\xi$ tends to
infinity, $A$ must be zero. This implies that $f_0$ is a function of $\hat\tau$ alone,
 and this yields some partial differential equation to be
satisfied by $\varphi_0$ (or $\varphi$). This equation is obtained from equation (\ref{23})
by simplifying the terms containing $f$. It is
\begin{equation}
\partial_{\xi_1}\varphi_0+\gamma\partial_\zeta\varphi_0
+\frac1{v_1}\partial_\tau\varphi_0-i\rho_1\varphi_0=0 . \label{26}
\end{equation}
The solution of equation (\ref{26}) is
\begin{equation}
\varphi_0=\varphi_1\left(\xi_2,\tau-\frac\xi v-\frac{\xi_1}{v_1}
,\eta,\zeta-\gamma\xi_1\right) \,e^{i\rho_1\xi_1} . \label{27}
\end{equation}
$1/{v_1}$ is the first correction to the inverse of the group velocity
\begin{equation}
\frac1{v_1}=k_{\frac12,o}^{\prime}\,.
\end{equation}
$\rho_1$ is the next correction to the wave vector $k$,
\begin{equation}
\rho_1=k_{\frac32,o}+\frac{\omega}{4nc}\Delta n_o^2(\omega)\gamma^2.
\label{28}
\end{equation}
The dependency relative to $(\zeta-\gamma\xi_1)$ does not represent a
walk-off: it simply gives account for wave propagation in a direction making
the small angle $\gamma\sqrt\varepsilon$ with the $x$-axis.

In an analogous way, the $\xi$-evolution of the second harmonic is found,
\begin{equation}
g=g_0\left(\tau-\frac\xi v\right) \,e^{i\delta_2\xi},  \label{29}
\end{equation}
with
\begin{equation}
\frac{1}{v_{1}}=k_{\frac{1}{2},e}^{\prime }\,,
\end{equation}
and
\begin{equation}
\rho _{2}=k_{\frac{3}{2},e}-\frac{\omega \gamma ^{2}}{2nc}\left( \Delta
n_{e}^{2}(2\omega )-2\Delta n_{o}^{2}(2\omega )\right) . \label{31}
\end{equation}
Equation (\ref{30}) is deduced from equation (\ref{29}). As
previously, the variables must be the same in expression
(\ref{27}) for $\varphi $, as in (\ref{30}) for $\psi $. In
particular, the correction term~$1/{v_{1}}$ to the inverse of the
group velocity must be the same in both cases. Thus
\begin{equation}
k_{\frac{1}{2},o}^{\prime }=k_{\frac{1}{2},e}^{\prime }.
\end{equation}

Other conditions give
\begin{gather}
\vec{E}_{3}^{1}=\left(
\begin{array}{c}
\frac{i}{k}\partial _{\eta }\varphi \\
F \\
0
\end{array}
\right),  \label{33}
\\
\vec{E}_{3}^{2}=\left(
\begin{array}{c}
-\gamma g+\frac{i}{2k}\partial _{\zeta }\psi +\frac{\gamma }{2n^{2}}\left[
2\Delta n_{o}^{2}(2\omega )-\Delta n_{e}^{2}(2\omega )\right] \psi \\
0 \\
G
\end{array}
\right)  .\label{34}
\end{gather}
$F$, $G$ are functions to be determined. The expressions of
$E_{3+\frac12}^{1,x}$, $E_{3+\frac12}^{2,x}$ are also obtained, they are use in the remainder
of the computation:
\begin{gather}
E_{3+\frac12}^{1,x}=\frac{i}k\partial_\eta f-\frac{\Delta n_o^2(\omega)}{2n^2}
 \partial_\eta \varphi,  \label{35}
\\
E_{3+\frac12}^{2,x}= -\gamma G+\frac{i}{2k}\partial_\zeta g
+\frac\gamma{2n^2}\left[2\Delta n_o^2-\Delta n_e^2\right] g \nonumber\\
\phantom{E_{3+\frac12}^{2,x}=} +\left[\frac{\Delta n_o^2}{2n^4}\left(\Delta n_e^2-2\Delta
n_o^2\right) +\frac1{n^2}\hat\chi_{1,o}^{(1)}-\gamma^2\right]\gamma\psi
\nonumber\\
\phantom{E_{3+\frac12}^{2,x}=}+\frac{i\gamma}{2k}\partial_\xi\psi+ \frac{i}{4kn^2}
\left(\Delta n_e^2-2\Delta n_o^2\right) \partial_\zeta \psi+\frac{i\gamma}{kv}
\partial_\tau\psi.
\label{36}
\end{gather}
(All indices in equation (\ref{36}) are taken at $2\omega$).

\subsection{The evolution equations}

At order $\varepsilon^4$, the evolution equations for $\varphi $ and $\psi$
are obtained as follows: the equation of order $\varepsilon^4$ for the $y$-component
of the fundamental is a partial differential equation involving $F $, $f$, and $\varphi$.
 Using the variables $\hat\xi$ and $\hat\tau$
defined by equation (\ref{24}), and writing $F=F_0(\hat\xi,\hat\tau)e^{i\delta_1\hat\xi}$,
it is found that $\partial_{\hat\xi}F_0$ is equal to some
expression depending on $\varphi_0$, $f_0$, and their derivatives. This
expression does not depend on $\hat\xi$. Thus, as shown above for $f_0$,
$F_0 $ is a function of $\hat\tau$ only, and the terms depending on $F$
cancel in the equation.

Then considering equation (\ref{27}), $f_{0}$ is
\begin{equation}
f_{0}=f_{1}\left( \xi _{1},\xi _{2},\tau -\frac{\xi _{2}}{v_{1}},\eta ,\zeta
-\gamma \xi _{1}\right) \,e^{i\rho _{1}\xi _{1}} . \label{32b}
\end{equation}
The notation $\partial _{\hat{\xi}_{1}}$ holds for the partial derivative
relative to $\xi _{1}$ in the variables defined by equation (\ref{32b}).
Transferring (\ref{27}) and (\ref{32b}) into the equation, the following result
is obtained: $\partial _{\hat{\xi}_{1}}f_{1}$ is equal to some expression
that depends on $\varphi _{1}$ only. Thus $\partial _{\hat{\xi}_{1}}f_{1}$
does not depend on $\xi _{1}$. Because $f_{1}$ must be bounded as $\xi _{1}$
tends to infinity, $f_{1}$ is independent of $\xi _{1}$ too. Then $f_{1}$
disappears from the equation, that reduces to the evolution equation (\ref
{33b}) for $\varphi $, taking into account the nonlinear term computed in
the following subsection.

The equation for the second harmonic is obtained in a similar way. It comes
from the equation of the multiscale expansion at order $\varepsilon ^{4}$,
for the $z$-component of the second harmonic. In the same way as for the
functions $F$ and $f$ in the case of the fundamental, it is seen that the
functions $G$ and $g$ must have the form
\begin{equation}
G=G_{0}\left( \tau -\frac{\xi }{v}\right) e^{i\delta _{2}\xi } , \label{39}
\end{equation}
and
\begin{equation}
g=g_{1}\left( \xi _{2},\tau -\frac{\xi }{v}-\frac{\xi _{1}}{v_{1}},\eta
,\zeta -\gamma \xi _{1}\right) e^{i(\delta _{2}\xi +\rho _{2}\xi _{1})}.
\label{40}
\end{equation}
Then all terms involving either $G$ or $g$ cancel in the equation, and the
evolution equation (\ref{41}) for $\psi $ is obtained.

\subsection{Computation of the nonlinear terms}

The nonlinear term in equation (\ref{33b}) for $\varphi $ is
\begin{equation}
-\frac{\omega ^{2}}{c^{2}}P_{4}^{1,y},
\end{equation}
where $P_{4}^{1,y}$ is the $y$-component in the coefficient $\vec{P}_{4}^{1}$
of $\varepsilon ^{4}e^{i\phi }$, in the expansion of the nonlinear
polarization $\chi ^{(2)}:\vec{E}\vec{E}$. It is straightforwardly seen that
\begin{equation}
\vec{P}_{4}^{1}=2\hat{\chi}^{(2)}(2\omega ,-\omega ):\vec{E}_{2}^{2}\vec{E}_{2}^{-1}  .\label{35b}
\end{equation}
The value of this nonlinear term depends on the symmetry
properties of the $\chi ^{(2)}$-tensor. For the $\bar{4}2{\rm m}$
class of crystals, to which belongs e.g.\  KDP, in the crystal
axes frame \cite{boyd}, the non-vanishing components of the $\chi
^{(2)}$-tensor are
\begin{equation}
\chi _{xyz}^{(2)}  =  \chi _{yxz}^{(2)},\qquad\chi _{xzy}^{(2)} =  \chi _{yzx}^{(2)},\qquad
\chi _{zxy}^{(2)} = \chi _{zyx}^{(2)}.
\label{36b}
\end{equation}
and all other $\chi ^{(2)}$-components are zero. Thus, if the axes of the
coordinates frame are the crystal axes, because $\vec{E}_{2}^{2}$ and $\vec{E}_{2}^{1}$
are parallel to the $z$- and to the $y$-axis respectively, $\vec{P}_{4}^{2,y}$
would be zero. Thus the crystal axes must be rotated for some
angle $\beta $ around the optical axis (the $z$-axis). We call $(x^{\prime },y^{\prime },z^{\prime })$
 this rotated crystal axes frame.
Relative to it, the nonlinear term becomes
\begin{equation}
-\frac{\omega ^{2}}{c^{2}}\vec{P}_{4}^{1}=A_{1}\left(
\begin{array}{c}
0 \\
0 \\
1
\end{array}
\right) \psi \varphi ^{*},  \label{37}
\end{equation}
with
\begin{equation}
A_{1}=\frac{2\omega ^{2}}{c^{2}}\sin 2\beta
\;\chi _{x^{\prime }y^{\prime}z^{\prime }}^{(2)}(2\omega ,-\omega ) . \label{38}
\end{equation}

For the 3m symmetry class, the nonzero components of the $\chi _{(2)}$-tensor are~\cite{boyd}
\begin{gather}
\chi _{xzx}^{(2)}=\chi _{yzy}^{(2)},\qquad\chi _{xxz}^{(2)}=\chi_{yyz}^{(2)},
\qquad\chi _{zxx}^{(2)}=\chi _{zyy}^{(2)},\qquad\chi _{zzz}^{(2)},\nonumber\\
\chi _{yyy}^{(2)}=-\chi _{yxx}^{(2)}=-\chi _{xxy}^{(2)}=-\chi _{xyx}^{(2)}.
\label{36l}
\end{gather}
The coefficient $A_{1}$ is independent from the angle $\beta $ introduced
for the ${\rm \bar{4}2m}$ class, and has the expression
\begin{equation}
A_{1}=-\frac{2\omega ^{2}}{c^{2}}\hat{\chi}_{zxx}^{(2)}(\omega ,\omega ).
\label{38l}
\end{equation}
Other values for this constant are obtained with other crystal symmetries.
They are easily computed, or can be found e.g.~in \cite{dmitriev}.

In equation (\ref{41}) for $\psi $, the nonlinear interaction constant $A_{2} $ is defined in the general case by
\begin{equation}
-\frac{4\omega ^{2}}{c^{2}}P_{4}^{2,z}=A_{2}\,\varphi ^{2} . \label{35t}
\end{equation}
For both the $\bar{4}2{\rm m}$ and the 3m symmetry classes, the use of the $\chi ^{(2)}$-structure (\ref{36b})
or (\ref{36l}), and the complete symmetry
property of the $\chi^{(2)}$-tensor, proves the equality $A_{2}=2A_{1}$.

This completes the derivation of the model equations (\ref{33b})--(\ref{41}).
\label{leblond-lastpage}

\end{document}